\RequirePackage{fix-cm}

\documentclass[smallextended]{svjour3} 
\smartqed  

\usepackage{amssymb,amsmath,amsfonts,latexsym} 
\usepackage{hyperref}
\usepackage{listings}

\usepackage{mathtools}
\usepackage{multirow,array}
\usepackage{graphicx}
\usepackage{subfigure}
\usepackage{epstopdf}
\usepackage{float} 
\usepackage{color, xcolor}
\usepackage{listings}
\usepackage{framed}
\newfloat{listing}{h}{listOfListings}[section]
\floatname{listing}{Listing}

\usepackage{xcolor}
\usepackage{listings}

\lstdefinestyle{mystyleMathematica134}{
    basicstyle=\footnotesize\ttfamily,
    keywordstyle=\bfseries,
    escapechar=ä,
    extendedchars=true,
    numbers=left,
    numberstyle=\tiny,
    xleftmargin=1.0em,
    stepnumber=1,
    numbersep=6pt,
    captionpos=b,
    breaklines=true,
    commentstyle=\color{gray!95!black},
    morecomment=[l][\color{gray!100!black}]{(*},
    morecomment=[s][\color{gray!100!black}]{*)},
    stringstyle=\ttfamily,
    showstringspaces=false,
    language=Mathematica,
    morekeywords={
      LinearSolve,
      Solve,
      ParallelTable,
      ListContourPlot3D,
      GraphicsGrid,
      FractionalD,
      CaputoD
    }
}

\lstdefinestyle{mystyleCpp}{
    basicstyle=\footnotesize\ttfamily,
    keywordstyle=\bfseries,
    extendedchars=true,
    numbers=left,
    numberstyle=\tiny,
    xleftmargin=1.0em,
    stepnumber=1,
    numbersep=6pt,
    captionpos=b,
    breaklines=true,
    commentstyle=\color{gray!90},
    stringstyle=\ttfamily,
    showstringspaces=false,
    language=C++,
    morekeywords={
      class,
      public,
      private,
      mlf,
      setN,
      FloatTypesAllowed,
      printList,
      polySum,
      horner,
      ghGamma,
      trapez,
      mlfPadeApprox,
      setOrder,
      mittagLefflerE
    }
}

\allowdisplaybreaks

\journalname{Fract. Calc. Appl. Anal.} 

\begin{document}


\title{Combining 
arbitrary order global Pad\'e approximation of the Mittag-Leffler function with its addition formula for a significant accuracy boost}

\titlerunning{Global Pad\'e approximation and addition formula}

\author{
    Richard Herrmann$^1$ 
}

\authorrunning{R. Herrmann} 

\institute{Richard Herrmann$^{1,*}$
\at
gigaHedron.de, Berliner Ring 80,  63303 Dreieich, Germany \\
\email{r.herrmann@fractionalcalculus.org} $^*$ corresponding author 
}

\date{Received: 1 September 2024 / Revised:  / Accepted: ......}


\maketitle

\begin{abstract}
The combination of the global Pad\'e approximation of the Mittag-Leffler function with its addition formula 
for the case $\alpha<1$ 
yields significantly higher accuracy results for a given arbitrary order $n$.  
We present a solution in terms of a Mathematica notebook to determine the general structure of the system of linear
equations to be solved, followed by an implementation as a {\tt{C++}} program using the {\tt{Eigen}} template library 
for linear algebra. 
For a comparison with contour integral solutions we  present an implementation as a {\tt{C++}} program using the
{\tt{boost}} library's quadrature package employing the Gauss-Kronrod-method. 


\end{abstract} 


%
%
\section{Introduction}
\setcounter{section}{1} \setcounter{equation}{0} 

Because of their fundamental importance within the framework of fractional calculus \cite{mai20}, 
especially in solving fractional differential equations \cite{gor02}, the Mittag-Leffler functions 
are essential for anyone working in this field. They are necessary in both verifying analytic results 
and exploring practical applications numerically.

Although known for more than 120 years, the Mittag-Leffler function was only recently included 
in the Mathematical Subject Classification (MSC) system as code 33E12, starting with the 2000 revision. 
Fortunately, the history numerical of implementation began much earlier:

Beginning with the direct truncated power series expansion, several methods for numerical evaluation have been used 
and combined to optimize the efficiency and accuracy of numerical implementation.

Widely used is the numerical evaluation of an integral representation along a Hankel contour
 \cite{gor97b}, \cite{hil06}, \cite{die05} applying standard quadrature methods.

Alternatively numerical inversion of the Laplace transform \cite{gar15}, \cite{ort19} is common practise. 
Furthermore, global Pad\'e approximation methods \cite{zen13}, \cite{sar20} are used too.

For large arguments there exist useful asymptotic expansions \cite{gor97b}. 

Nevertheless, every method has its limitations, making it tempting to use library implementations, which have become
available recently for CAS Matlab (since 2006), Mathematica (since 2012),  Python (mpmath since 2010, SciPy since 2017)  and of course, 
third-party implementations for almost all programming languages. But they are very slow. 

In the following, we will develop a {\tt{C++}} implementation of a symmetric order $n, n$  case of the global Pad\'e approximant, based on the
method proposed by \cite{zen13} and refined by \cite{sar20} which will improve the accuracy and performance significantly.

We will make the following steps:
\begin{itemize}
\item Present two analytic approaches to derive the global Pad\'e approximant of the Mittag-Leffler function, first explicitly and 
then only the system of linear equations in terms of a Mathematica notebook.  
\item Solve the derived system of linear equations using the Eigen linear algebra package within a {\tt{C++}} program.
\item Compare the performance and stability to a classical quadrature algorithm for a solution of the integral representation
of the Mittag-Leffler function using the {\tt{boost}} quadrature package. 
\item From an error analysis of compared strategies  we make suggestions to significantly enhance the precision ond performance
of our approximation.  
\end{itemize}

\section{Setting up the algebra }
A classical approach to  approximate a function is a Taylor expansion eg. at the origin . For  the Mittag-Leffler function we may use the defining series directly:
\begin{eqnarray}
\label{num203001start}
E_{\alpha,\beta}(z) &=&  \sum_{k=0}^{\infty}\frac{z^k}{\Gamma(\beta+\alpha k)}
\end{eqnarray}
We will  give a global Pad\'e approximant of order $n$ 
\begin{equation}
\label{hg02aPA}
{_n}P_{n+1}(\{a_i\};\{b_j\};x) = \frac{p(x)}{q(x)} =
\frac{  \sum_{i=0}^{n} a_i x^i}
{ \sum_{j=0}^{n+1} b_j x^j
}   
\end{equation}
of the Mittag-Leffler function $E_{\alpha,\beta}(-x)$, $x \in \mathbb{R}$
which covers all $x$ in the region  $0 \leq x < \infty$ in the range $0 \leq \alpha \leq 1$ and $0 \leq \beta \leq 2$. 

In \cite{he18c}, we determined  $2 n$ coefficients for the global Pad\'e approximant by  the requirement, that $2 n$  derivatives of the Mittag-Leffler function and its Pad\'e approximant exactly agree  at
the origin $x=0$. Talking about boundary conditions for differential equations, this strategy could be compared
with the model of a vibrating rod, fixed at the left end, but otherwise freely moving.

Within this picture, we could alternatively consider the model af a vibrating string, which is clamped at both ends.
This is the basic idea proposed in \cite{zen13}:

Use the asymptotic expansions of order $n$ of the Mittag-Leffler function 
for both cases simultaneously $\lim_{x \rightarrow 0} $ and $\lim_{x \rightarrow \infty}$.  

For the case $0<\alpha<1$, it follows from the definition (\ref{num203001start}):
\begin{eqnarray}
\label{num203001a}
\lim_{x \rightarrow 0 } \Gamma(\beta-\alpha) x E_{\alpha,\beta}(-x) &\approx&
\Gamma(\beta-\alpha) x \sum_{k=0}^{n-1}\frac{(-x)^k}{\Gamma(\beta+\alpha k)} \\
&\equiv&  a(x)  + O(x^{n+1})\\
\label{num203001b}
\lim_{x \rightarrow \infty }  \Gamma(\beta-\alpha) x E_{\alpha,\beta}(-x)&\approx&
-\Gamma(\beta-\alpha) x \sum_{k=1}^{n}\frac{(-x)^{-k}}{\Gamma(\beta-\alpha k)}\\
&\equiv& b(1/x) + O(x^{-n})
\end{eqnarray}
at $x = 0$ and $x = +\infty$, respectively. 

Now we determine the $2 n$ coefficients for the global Pad\'e approximant  by the requirement,
that $n$ terms in both  the  Pad\'e approximant and Mittag-Leffler function $E_{\alpha,\beta}(-x)$ should exactly agree 
at $x=0$ and $1/x=0$ simultaneously in the series $p(x)$ and $q(x)$ from (\ref{hg02aPA})
compared with $a(x)$ in (\ref{num203001a}) and $b(x)$ in (\ref{num203001a}) for all terms with given 
exponent $m= \{0,1,2,...,n-1\}$ in $x$.  
\begin{eqnarray}
\label{hg03abp}
p(x)  &=& q(x) a(x)               \qquad\qquad  \forall m = 0,1,2,...,n \\
p(x) x^{-n} &=& q(x) x^{-n} b(x)  \qquad\,\,  \forall m = 1,2,...,n-1
\end{eqnarray}

Finally $p_n(x) = 1, q_n(x) = 1$ guarantees the correct asymptotic behaviour of the approximation \cite{zen13}. 

E.g. for the case $n=2$ we have from (\ref{hg02aPA})
\begin{eqnarray}
\label{hg03n2}
\Gamma(\beta- \alpha) x E_{\alpha, \beta}(x)  &\approx& 
{p_0 + p_1 x + x^2 \over q_0 + q_1 x + x^2}
\end{eqnarray}
with a set of unknown variables $\{p_0,p_1,q_0,q_1\}$.

According to (\ref{hg03abp}) we obtain an inhomogeneous system of $3+1=4$ equations:
\begin{eqnarray}
\label{hg034}
p_0  &=&  0 \\
p_1 - q_0{  \Gamma(\beta-\alpha) \over \Gamma(\beta)} &=&  0 \\
1   - q_1{  \Gamma(\beta-\alpha) \over \Gamma(\beta)} + q_0{\Gamma(\beta-\alpha) \over \Gamma(\alpha+\beta)}&=&  0 \\
p_1 - q_1 + { \Gamma(\beta-\alpha) \over \Gamma(\beta- 2 \alpha)} &=&  0 
\end{eqnarray}
Mathematica provides several methods to solve such sets of equations. We first choose the simplest way using {\tt{Solve}}:

\begin{figure}[t]
\begin{center}
\includegraphics[ width=104mm]{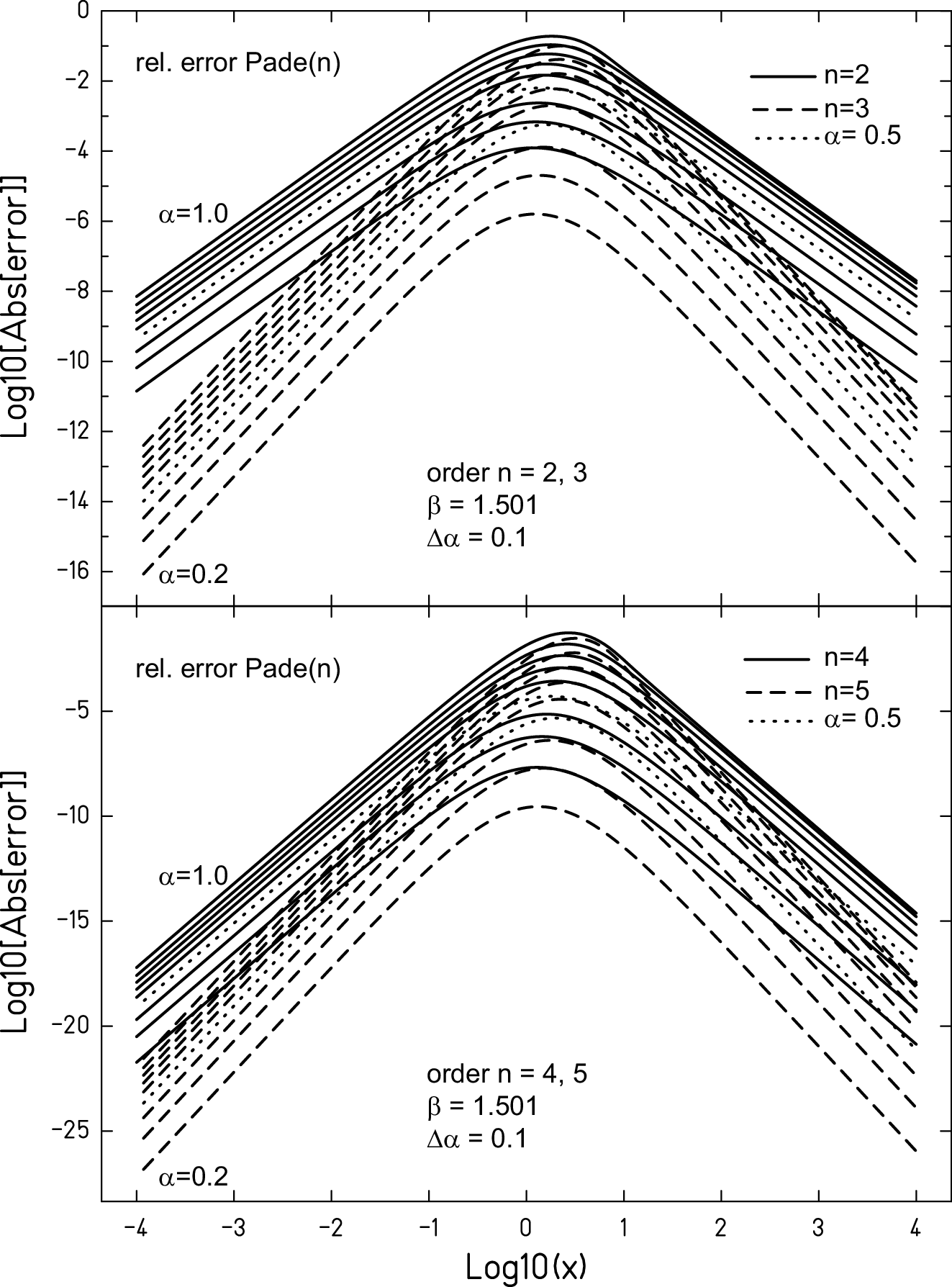}\\
\caption{
\label{bnum03_31}
{ Top: Double logarithmic plot of the relative error $\rho(x)$ within the range $10^{-4} \leq x \leq 10^4$ for 
the Pad\'e approximants of order $n=2$ (thick lines) 
  and $n=3$ (dashed lines) of the generalized Mittag-Leffler function $E_{\alpha,\beta}(-x)$ for
  $\beta = 1501/1000$ and  $2/10 \leq \alpha \leq 1 $ in $\Delta \alpha = 1/10 $ steps. For $\alpha = 5/10$ we
  used pointed lines. Within the project rational numbers have been used for arbitrary precision calculus.
  Bottom: Same, but orders order $n=4$ (thick lines) 
  and $n=5$ (dashed lines). 
} }
\end{center}
\end{figure}

Listing \ref{bnum03_p1} presents this solution extending the work of \cite{zen13} from order $n=2$ to arbitrary order $n$. 
\begin{listing}[t]
\begin{oframed}
\vskip-3mm
\begin{lstlisting}[
  label={bnum03_p1},
  style=mystyleMathematica134,
  caption={Listing of pade1.nb}
]
                    (* order of  PadeApproximant *)
n = 2                 
                                  (* asymptotics *)
a[x_]:= +Gamma[bet-alp] x *
    Sum[(-x)^k/Gamma[bet + alp k], {k,0,n+1}];
b[x_]:= -Gamma[bet-alp] x *
    Sum[(-x)^(-k)/Gamma[bet - alp k], {k, 1, n}];
                           (* collect parameters *)
ps = Symbol["p" <> ToString[#]] & /@ Range[0, n - 1];
qs = Symbol["q" <> ToString[#]] & /@ Range[0, n - 1];
vars = Append[ps, qs] // Flatten;
                                  (* powers of x *)
xs = (x^#) & /@ Range[0, n - 1];
                         (* Pade polynomials p/q *)
p = Total[ps xs] + x^n; 
q = Total[qs xs] + x^n;
                           (* collect conditions *)
ca = CoefficientList[p - q a[x], x];
cb = CoefficientList[p/x^n - q/ x^n b[x], 1/x];
                            (* collect equations *)
condsa = (ca[[#]] == 0) & /@ Range[n + 1];
condsb = (cb[[#]] == 0) & /@ Range[2, n];
conds = Append[condsa, condsb] // Flatten
(* solve system of inhomogenous linear equations *)
result = Solve[conds, vars];
                             (* present solution *)
padeApproximant = p/q/x/Gamma[bet - alp] /. result 
\end{lstlisting}
\vskip-3mm
\end{oframed}
\end{listing}
\index{listings!Pad\'e approximant 1}

\begin{itemize}
\item \ref{bnum03_p1}:1--2: Input parameter order $n$.
\item \ref{bnum03_p1}:3--7: Asymptotic expansions for $a(x)$ and $b(x)$.
\item \ref{bnum03_p1}:8--11: Generate the list of all unknown parameters.
\item \ref{bnum03_p1}:12--13: Generate the list of all powers $\{x^0,x^1,...x^{n-1}\}$.
\item \ref{bnum03_p1}:15--16: Generate the polynomials $p(x)$ and $q(x)$.
\item \ref{bnum03_p1}:18--23: Generate the system of equations (\ref{hg03abp}).
\item \ref{bnum03_p1}:24--27: Apply the Mathematica method {\tt{Solve}} to obtain the parameter values and insert into the padeApproximant.
\end{itemize}

In Figure \ref{bnum03_31} we compare the accuracy of the Pad\'e approximation method from Listing \ref{bnum03_p1} 
printing the relative error $\rho_n(\alpha, \beta, x)$
\%
\begin{equation}
\label{bnum3rho}
\rho_n(\alpha, \beta, x) = {E_{\alpha,\beta}(x) - P_n(\alpha, \beta,x) \over E_{\alpha,\beta}(x)}
\end{equation}
with
the exact values for the Mittag-Leffler function $E_{\alpha, \beta}(-x)$ within the range  $10^{-4} \leq x \leq 10^4$ 
for orders $n=\{2,3\}$ and $0.2 \leq \alpha \leq 1.0$ for a fixed $\beta = 1.501$. 

Actually we wanted to use $\beta = 3/2$ but according to (\ref{num203001b}) we have terms of type 
$1 / \Gamma(\beta-\alpha k)$ which look harmless but for $n \geq 2$ for all  $\beta-\alpha k = 0, -1,-2, ...$ 
we are in trouble. 
A classical example for agile software development. 

But this is only the first problem of this approach.

Let us have a look at the error behaviour of the Pad\'e approximant for increasing order $n$. 
As we can deduce from Figure \ref{bnum03_31}, stepping from order $n=2$ to $n=3$:

For small $\alpha$
we of obtain an accuracy enhancement  for e.g. $\alpha=0.2$ of a factor $\approx 80 $ or 2 orders of magnitude (
from 
$\max(\log_{10}\rho_2(0.2, 1.501, x))\approx -3.9 $, 
$\max(\log_{10}\rho_3(0.2, 1.501, x))\approx -5.8 $,  
$\max(\log_{10}\rho_4(0.2, 1.501, x))\approx -7.7 $,
$\max(\log_{10}\rho_5(0.2, 1.501, x))\approx -9.6 $
). 

But near  $\alpha =1$ we have   
e.g. increasing the order for $\alpha=0.9$ a factor of only $\approx 2 $ or $1/5$ orders of magnitude (
from 
$\max(\log_{10}\rho_2(0.9, 1.501, x))\approx -0.72 $,
$\max(\log_{10}\rho_3(0.9, 1.501, x))\approx -0.99 $, 
$\max(\log_{10}\rho_4(0.9, 1.501, x))\approx -1.26 $,

\noindent
$\max(\log_{10}\rho_5(0.9, 1.501, x))\approx -1.52 $
).

This general trend may be summarized: convergence is fast for small $\alpha$ and becomes weaker for large
$\alpha$. This supports the findings of \cite{sar20}, who for $n=4$ report similar maximum relative errors.

Assuming a logarithmic behaviour of the error $O(\log(n))$ we have to increase the order $n$ for a given
accuracy. 
In our example ($\beta \approx 1.5$), 
if an accuracy  of about 8 digits is required, this is already almost fulfilled 
for $\alpha \leq 0.2$ setting $n \geq 4 $.
For $\alpha \geq 0.9$ we roughly estimate $n \geq 39$.   

At this stage a third point plays an increasingly important role. The time needed to calculate the exact general
solution for the coefficients of the Pad\'e approximant increases roughly as $O(n^3)$ but even worse, there is
no significant simplification or length reduction for the exact formula.

Therefore in a next step, we decided to use Mathematica  to generate only the linear system of equations
but the solution itself should be with a linear algebra package in {\tt{C++}}.

\section{Setting up the system of linear equations}

\begin{listing}[t]
\begin{oframed}
\vskip-3mm
\begin{lstlisting}[
  label={bnum03_p2},
  style=mystyleMathematica134,
  firstnumber=28,
  caption={Listing of pade2.nb}
]
Print["setup matrix for algebra package solver m.x + bs"];
vector = 
Append[ca[[#]] &/@ Range[m], cb[[#]] &/@ Range[2, n]] 
// Flatten;
coefficientsMatrix = Coefficient[#, vars] & /@ vector ;
bs = coefficientsMatrix . vars - vector;
coefficientsMatrix /= Gamma[bet - alp];
bs /= Gamma[bet - alp];
                           (* show traditional form *)
coefficientsMatrix // MatrixForm // TraditionalForm  
bs                               // TraditionalForm
(* e.g. solve system of inhomogenous linear equations *)
alp = 0.5;                                     (* alpha *)
bet = 1.501;                                   (* beta  *)
res = FullSimplify[LinearSolve[coefficientsMatrix, bs]]
\end{lstlisting}
\vskip-3mm
\end{oframed}
\end{listing}
\index{listings!Pad\'e approximant 2}

For that purpose we will add a piece of code to our Listing \ref{bnum03_p1}. 

This should transfer 
the system of conditions $\{ ca, cb \}$ given in \ref{bnum03_p1}:18--19 to the standard form of a 
linear inhomogeneous system of equations:
\begin{eqnarray}
m \vec{x} & =& \vec{b} 
\end{eqnarray}
with the $2 n \times 2 n$ coefficient  matrix $m$, $2 n$ unknown variables $\vec{x}$ 
and the $2 n$ dimensional  inhomogeneity vector $\vec{b}$.
In Listing \ref{bnum03_p2}  we present the realization.  

\begin{itemize}
\item \ref{bnum03_p2}:28--31: A help vector is constructed from the conditions $\{ ca, cb \}$.
\item \ref{bnum03_p2}:32--33: The coefficient matrix $m$ and the vector $b$(s) are generated.
\item \ref{bnum03_p2}:34--35: The common factor $\Gamma(\beta - \alpha)$ is eliminated from $m$ and $b$(s).
\item \ref{bnum03_p2}:37--38: View the result in traditional form
\item \ref{bnum03_p2}:39--42: Quality control:  the linear system is solved using the Mathematica function 
{\tt{LinearSolve}}. Should yield the same value as using {\tt{Solve}} in Listing \ref{bnum03_p1}.  
\item \ref{bnum03_p1}:18--23: Generate the system of equations (\ref{hg03abp}).
\item \ref{bnum03_p1}:24--27: Apply the Mathematica method {\tt{Solve}} to obtain the parameter values and insert into the Pad\'e Approximant.
\end{itemize}

As an example, for $n=4$ we obtain (note the slight difference to  \cite{sar20}):
\begin{eqnarray}
\label{bnum3_equs}
m \vec{x} & =& \vec{b}  \qquad \qquad n=4
\end{eqnarray}
with
\begin{eqnarray}
\label{bnum03_m}
m  & =&  
\left(
\begin{array}{cccccccc}
1 & 0 & 0 & 0 & 0 & 0 & 0 & 0 \\
0 & 1 & 0 & 0 & -\frac{\Gamma
(\beta-\alpha)}{\Gamma (\beta)} & 0 & 0 & 0
\\
0 & 0 & 1 & 0 & \frac{\Gamma
(\beta-\alpha)}{\Gamma (\alpha+\beta)}
& -\frac{\Gamma (\beta-\alpha)}{\Gamma
(\beta)} & 0 & 0 \\
0 & 0 & 0 & 1 & -\frac{\Gamma
(\beta-\alpha)}{\Gamma (2
\alpha+\beta)} & \frac{\Gamma
(\beta-\alpha)}{\Gamma (\alpha+\beta)}
& -\frac{\Gamma (\beta-\alpha)}{\Gamma
(\beta)} & 0 \\
0 & 0 & 0 & 0 & \frac{\Gamma
(\beta-\alpha)}{\Gamma (3
\alpha+\beta)} & -\frac{\Gamma
(\beta-\alpha)}{\Gamma (2
\alpha+\beta)} & \frac{\Gamma
(\beta-\alpha)}{\Gamma (\alpha+\beta)}
& -\frac{\Gamma (\beta-\alpha)}{\Gamma
(\beta)} \\
0 & 0 & 0 & 1 & 0 & 0 & 0 & -1 \\
0 & 0 & 1 & 0 & 0 & 0 & -1 & \frac{\Gamma
(\beta-\alpha)}{\Gamma (\beta-2
\alpha)} \\
0 & 1 & 0 & 0 & 0 & -1 & \frac{\Gamma
(\beta-\alpha)}{\Gamma (\beta-2
\alpha)} & -\frac{\Gamma
(\beta-\alpha)}{\Gamma (\beta-3
\alpha)} \\
\end{array}
\right) \\
\label{bnum03_b}
b^T  & =&  
\left(
0,0,0,0,-1,-\frac{\Gamma
(\beta-\alpha)}{\Gamma (\beta-2
\alpha)},\frac{\Gamma (\beta-\alpha)}{\Gamma
(\beta-3 \alpha)},-\frac{\Gamma
(\beta-\alpha)}{\Gamma (\beta-4
\alpha)}
\right) \\
\label{bnum03_x}
x^T  & =&  
\left(
p_0,p_1,p_2,p_3,
q_0,q_1,q_2,q_3
\right)
\end{eqnarray}

Note the substructure quadrants in $m$ and the simple coefficient values based on the set of 
$\Gamma(\beta + k \alpha)$ with $k \in \{-n, ..., 0 , ... , n-1 \}$. Note also, that the terms of the type
$1 / \Gamma(\beta - k \alpha)$ in (\ref{bnum03_b}
) may be easily set to 0 when $\beta - k \alpha = 0,-1,-2,...$. Experimenting with
different orders $n$ reveals the general structure of this set of equations.  

We decided to employ the Eigen {\tt{C++}} template library for linear algebra \cite{eig21} to solve this set of equations
numerically.

For practical purpose we encapsulate the behaviour within a class 

\noindent
{\tt{GlobalPadeMLF}} and create a test routine  
{\tt{testGlobalPadeMLFfirst}}, to demonstrates the usage. 

The idea is to check the behaviour of the Pad\'e approximant 
for a given set of variables $\alpha, \beta, x$ with increasing order $n$, but of course, the tests may be more sophisticated 
in a later stage of development.

\begin{listing}
\begin{oframed}
\vskip-3mm
\begin{lstlisting}[
  label={bnum03_mlffirst},
  style=mystyleCpp,
  language=C++,
  firstnumber=1,
  caption={Test of class GlobalPadeMLF}
]
#include <iostream>
#include "globalPadeMLF.hpp"

using namespace std;

int main() {
// create an instance of the class GlobalPadeMLF
GlobalPadeMLF mlfPadeApprox;  // constructor

double alpha{0.9};
double beta{1.5};
double x{-1};
// exact value of E_{9/10, 3/2}(-1) from CAS
double exact{0.59595802527072791093339988837073}; 

// print title and column headers
cout << "E("<<alpha<<","<<beta<<","<<x<<")" << endl;
cout << "n log10(abs(err))" << endl;

// print error for increasing orders n=2...15
for (int n = 2; n < 15; n++) {
mlfPadeApprox.setOrder(n);  
double approx = mlfPadeApprox.mittagLefflerE(alpha, beta, -1.0); 
double relErr = ( approx - exact ) / exact; 
cout << n << " " << log10(abs(relErr)) << endl; 
}

return 0;
}    
\end{lstlisting}
\vskip-3mm
\end{oframed}
\end{listing}
\index{listings!test of Mittag Leffler contour integral }

This test routine is given in Listing \ref{bnum03_mlffirst} and performs the following steps:   
\begin{itemize}
\item \ref{bnum03_mlffirst}:2: Include the class {\tt{GlobalPadeMLF}} . 
\item \ref{bnum03_mlffirst}:8: Instance of class  {\tt{GlobalPadeMLF}} is created. 
\item \ref{bnum03_mlffirst}:10--14: Parameters for the test case $E_{0.9,1.5}(-1)$ are given.
\item \ref{bnum03_mlffirst}:21--28: For increasing order $2 \leq n \leq 14$ the relative error is calculated. using
the public methods {\tt{GlobalPadeMLF::setOrder}} and 

\noindent
{\tt{GlobalPadeMLF::mittagLefflerE}}.
\end{itemize}
\begin{listing}
\begin{oframed}
\vskip-3mm
\begin{lstlisting}[
  label={bnum02_1304},
  style=mystyleCpp,
  language=C++,
  firstnumber=1,
  caption={Methods of class GlobalPadeMLF}
]
class GlobalPadeMLF {
private: 
double alpha;           // current alpha
double beta;            // current beta
double alphaTresh{1.0}; // threshold addition formula
int n ;         // order of Pade Approximant = P(n)/Q(n)                       
int n2;         // total number of coefficients = 2*n
// Linear system  m . x = b  solve with Eigen library 
// m -> matM, x -> vecPsQs , b -> vecB
// 1/Gamma[beta + i*alpha] elements for matM, vecB
Eigen::VectorXd lstGammasP; 
// 1/Gamma[beta - i*alpha] elements for matM, vecB 
Eigen::VectorXd lstGammasM; 
Eigen::MatrixXd matM;        // matM * vecPsQs = vecB
Eigen::VectorXd vecB;        // inhomogeneity b    
Eigen::VectorXd vecPsQs;     // solution vector {Pn, Qn} 
void setLstGammas();         // create list of gammas
void setVectorB();           // set up vecB
void setMatrixM();           // set up matM 
// set alpha/beta dependent quantities and solve the
// system of linear equations for Pade-coefficients
void SetAlphaBetaSolveEqsGetPsQs(
double alpha_new, double beta_new = 1);
template < typename T >      // Evaluate Pade P(n)/Q(n)
T getPadeApproximant(T x);
template < typename T >      // a==b?true:false
bool almostEqual(T a, T b);
public:
GlobalPadeMLF();             // constructor
// set order n in PadeApproximant 
void setOrder(int new_n);   
// call MittagLeffler function 
double mittagLefflerE(double alpha_new, 
        double beta_new , double x );
}; // end class
\end{lstlisting}
\vskip-3mm
\end{oframed}
\end{listing}
\index{listings!Declarations in class  GlobalPadeMLF}

In Listing \ref{bnum02_1304} the declarations of the class {\tt{GlobalPadeMLF}} are given:
\begin{itemize}
\item \ref{bnum02_1304}:3--6: In the private section we have first listed the 
relevant state variables of the class: $\alpha,\beta$ and the order $n$ and derived quantity $n2 = 2 n$. 
For $\alpha_\text{thresh}>\alpha$ the addition formula (\ref{bnum03_addTalpha}) will be implemented.  
\item 
\ref{bnum02_1304}:10--15: state variables with data type from the {\tt{Eigen}} library:
\item
\ref{bnum02_1304}:10: {\tt{lstGammasP}}, list of $\Gamma$-values  with argument
\begin{equation}
\Gamma^+_k(x)  =  {1 \over \Gamma(\beta + k \alpha)  } \qquad k= 0,..., n
\end{equation}
\ref{bnum02_1304}:12: {\tt{lstGammasM}}, list of $\Gamma$-values  with argument
\begin{equation}
\label{bnum203_gammas}
\Gamma^-_k(x)  =
\begin{cases}
  0         &       \text{$\beta - k \alpha = 0$}\cr
  \{  {1 \over \Gamma(\beta - k \alpha)  } \}       &       \text{else} 
  \end{cases}
 \qquad k= 0,..., n
\end{equation}
\item 
\ref{bnum02_1304}:13: {\tt{matM}}, matrix $m$ from (\ref{bnum03_m})
\item 
\ref{bnum02_1304}:14: {\tt{matB}}, vector $b$ from (\ref{bnum03_b})
\item 
\ref{bnum02_1304}:15: {\tt{vecPsQs}}, vector $x$ from (\ref{bnum03_x}) holding the set of $\{P_n,Q_n\}$ coefficients of the
Pad\'e approximant. 
\item \ref{bnum02_1304}:16--18: Corresponding methods to set up $\Gamma$-function values and then vector $b$ and matrix $m$.
\item 
\ref{bnum02_1304}:21--22: method {\tt{SetAlphaBetaSolveEqsGetPsQs}} calculates for a given set of $\alpha,\beta$ the 
solutions $\{P_n,Q_n\}$ and stores them in {\tt{vecPsQs}}.
\item \ref{bnum02_1304}:23--26: Finally we have two methods {\tt{getPadeApproximant}} evaluating (\ref{hg02aPA}) and 
{\tt{almostEqual}} used for testing $a==b$. Both methods are realized as templates, since they are intended to be used for
any float being real or complex valued. 
\item \ref{bnum02_1304}:28--33: In the public section we have three methods the constructor {\tt{GlobalPadeMLF}}, 
{\tt{setOrder}} to vary the order $n$ at runtime and last not least  {\tt{mittagLefflerE}}, which returns the Mittag-Leffler function
value. An example for the usage of these public member functions is given in Listing \ref{bnum03_mlffirst}.
\end{itemize}
\begin{listing}
\begin{oframed}
\vskip-3mm
\begin{lstlisting}[
  label={bnum03_publicFirst},
  style=mystyleCpp,
  language=C++,
  firstnumber=1,
  caption={Public methods of class GlobalPadeMLF}
]
// public methods
// constructor
GlobalPadeMLF(): n(0), n2(0), alpha(-1), beta(-1) {
setOrder(2); 
}
// initialize dimensions of lists and vals
inline void setOrder(int new_n) {
if (new_n != n) {  // // only change
n = new_n;
n2 = 2 * n;
matM.resize(n2, n2);   // initialize m 
matM.setZero();
vecB.resize(n2);       // initialize b
vecB.setZero();
vecPsQs.resize(n2);    // initialize x
lstGammasM.resize(n+1);// initialize lists
lstGammasP.resize(n+1);
alpha = -1; // force triggers a change
beta = -1;  // force triggers a change
}
}
// calculate E_{\alpha, \beta}(x)
inline double mittagLefflerE( double alpha_new, 
          double beta_new, double x ) {
// ToDo: special cases e.g.  
if ( alpha_new > 1.0 ) {  // no addition formula yet 
cout<< "alpha > 1, not yet implemented " <<endl;
return numeric_limits<double>::quiet_NaN();  
}
x = -x;   // internally with f(x) = E(-x)
// main job: get the {p_n,q_n} coefficients 
SetAlphaBetaSolveEqsGetPsQs(alpha_new, beta_new);
return getPadeApproximant(x,n,n); // get result
}
\end{lstlisting}
\vskip-3mm
\end{oframed}
\end{listing}
\index{listings!Public members in GlobalPadeMLF.cpp}

\begin{listing}
\begin{oframed}
\vskip-3mm
\begin{lstlisting}[
  label={bnum03_private1},
  style=mystyleCpp,
  language=C++,
  firstnumber=1,
  caption={Private methods of class GlobalPadeMLF}
]
inline void setLstGammas() { // create list of gammas
double value;
for (int i = 0; i <= n; i++) {
value = i * alpha - beta;                   // -i
lstGammasM(i) =  (almostEqual(floor(value), value)) ? 
                       0 : 1 / tgamma(-value); // 
lstGammasP(i) = 1 / tgamma(beta + i*alpha); // +i
}
}
inline void setVectorB() {  // setup vecB
for (int i = 0; i < n; i++)
vecB(n+i) = (((n + i)%2) ? 1 : -1) * lstGammasM(i+1);
vecB(n) *= -1;
}
inline void setMatrixM() { // set up the matrix of size n
// upper left quadrant
for (int i=0; i<n; i++) matM(i, i) = lstGammasM(1);
// lower left quadrant
for (int i=1; i<n; i++) matM(i+n, n-i) = lstGammasM(1);
// upper right quadrant
for (int i=1; i <= n; i++)
for (int j=0; j<n-i+1; j++)
  matM(i+j, n-1+i) = ((j%2)? 1 :-1)*lstGammasP(j);
// lower right quadrant
for (int i=1; i<n; i++)
for (int j=0; j<n-i; j++)
  matM(n+i+j, n2-i) = ((j%2)? 1 :-1)*lstGammasM(j+1);
}
// solve eqs. system  for Pade approximation
inline void SetAlphaBetaSolveEqsGetPsQs(double alpha_new, double beta_new = 1) {
if (!almostEqual(alpha_new, alpha) 
|| !almostEqual(beta_new, beta)) {
alpha = alpha_new;
beta = beta_new;
setLstGammas();                // refresh gammas
setMatrixM();                  // now populate matrix
setVectorB();                  // now populate vector
vecPsQs = matM.fullPivLu().solve(vecB); // solve system 
vecPsQs = (((n) % 2) ? +1 : -1)*vecPsQs; // odd/even n
}
}
\end{lstlisting}
\vskip-3mm
\end{oframed}
\end{listing}
\index{listings!Private members in GlobalPadeMLF.cpp}

Besides the constructor there are only two public member functions. There explicit form is given in Listing  
\ref{bnum03_publicFirst}:
\begin{itemize}
\item
\ref{bnum03_publicFirst}:2--5: In the constructor we set all variables to initial values and set the order
of the approximation to 2. 
\item
\ref{bnum03_publicFirst}:6--27: The member function {\tt{setOrder}} reacts on a  change in order $n$. 
It resizes all elements with a $n$ dependence and forces $\alpha = \beta = -1 $ to trigger a change in these 
parameters. 
\item
\ref{bnum03_publicFirst}:27--42: The member function {\tt{mittagLefflerE}} performs the following actions:
Checking the validity of the input parameters: This is in general a large variety of conditions, nomely
are the input parameters within allowed bounds or special cases of the general case  e.g. $E_{1,1}(x) = e^x$.
We give only one example in lines 32--34 for the case $\alpha>1$, since we have not yet implemented 
the addition formula (\ref{bnum03_addTalpha}). 

In line 37 we change the sign of x, since internally we give an approximation of $E_{\alpha,\beta}(-x)$.
The main job is done calling the private member function {\tt{SetAlphaBetaSolveEqsGetPsQs}}, which for a given set 
of $\{\alpha, \beta\}$ calculates the $\{p, q\}$ coefficients for the nominator and denominator of the Pad\'e approximant.

Since these values do not change any more for varying x, in line 41 calculating the result is separated in the private method
{\tt{getPadeApproximant}}.   
\end{itemize}
The private members of class {\tt{GlobalPadeMLF}} prepare the inhomogenous system of equations (\ref{bnum3_equs}) appropriate for a solution with 
your preferred linear algebra library package. We present the setup for the {{Eigen}} template library in Listing \ref{bnum03_private1}:
\begin{itemize}
\item
\ref{bnum03_private1}:1--9: Population of the lists  {\tt{lstGammasM}} and {\tt{lstGammasP}} from (\ref{bnum203_gammas}).
The term {\tt{( almostEqual ( floor ( value ), value ))}} checks for negative integers.
\item
\ref{bnum03_private1}:10--13: Population of vector $b$, compare with (\ref{bnum03_b}).
\ref{bnum03_private1}:15--28: Population of matrix $m$, compare with (\ref{bnum03_m}).
\item
\ref{bnum03_private1}:29--40: Preparing $\{ m,b \}$ and solving the inhomogenous system of equations using methods of the
{\tt{Eigen}} template library.
in two steps:
LU decomposition of matrix m with complete pivoting followed  by solving the resulting equation.  
\end{itemize}
\noindent
In Listing \ref{bnum03_private2} we present a possible implementation for {\tt{getPadeApproximant}}. 

Since we will extend our presentation from
real to complex arguments when implementing the addition formula and since the Pad\'e approximant works for both data types we suggest the 
realization as  a template. 


\begin{itemize}
\item
\ref{bnum03_private2}:2--15: {\tt{getPadeApproximant}} combines two Horner schemes for the nominator and denominator
of the Pad\'e approximant. Since we will use float as well as complex arguments (see addition formula) we implement this
method as a template.
\item
\ref{bnum03_private2}:17--23: {\tt{almostEqual}} proves $ a==b $.
\end{itemize}

\begin{listing}
\begin{oframed}
\vskip-3mm
\begin{lstlisting}[
  label={bnum03_private2},
  style=mystyleCpp,
  language=C++,
  firstnumber=1,
  caption={Private methods / templates  of class GlobalPadeMLF II}
]
 // Evaluate the Pade Approximant using Horner scheme
template < typename T >
inline T getPadeApproximant(T x, int n1 , int n2 ) {
T nominator = 1.0;
T deNominator = 1.0;
int i;
 // for first n1 -> P(n1)
for (i = n1 - 1; i > 0; i--)      
nominator = nominator * x + vecPsQs[i];
 // for next  n2 -> Q(n2)
for (i =  n1 + n2 - 1; i >= n1 ; i--)    
deNominator = deNominator * x + vecPsQs[i];
 // P/Q
return ( nominator / deNominator * lstGammasM[1]); 
}
  // a==b for double / float / complex
template < typename T >
inline bool almostEqual(T a, T b) {
return (a == 0.0 || b == 0.0) ?
(abs(a-b)<=std::numeric_limits<T>::epsilon()) :
((abs(a-b)<=std::numeric_limits<T>::epsilon()*abs(a)) ||
(abs(a-b)<=std::numeric_limits<T>::epsilon()*abs(b)));
}
\end{lstlisting}
\vskip-3mm
\end{oframed}
\end{listing}
\index{listings!Private members/templates in GlobalPadeMLF.cpp}
\section{Implementing the addition formula and results}
Now we are ready to investigate the properties of the global Pad\'e approximant for arbitrary order $n$. 
In case, we are interested only in the $x$ dependence of the Mittag-Leffler function for given order $n$ and fixed $\alpha$ and $\beta$,
which is the case generating the figures \ref{bnum03_31} or solving a fractional differential equation with constant $\alpha$, the
coefficients in $\{ P,Q\}$ do not change any more and the problem is reduced to evaluate two polynomials with constant 
coefficients. 

We can reproduce results from Figure \ref{bnum03_31} and those presented in the literature (see Table \ref{bnum203tab1}).
We may deduce, that the error behaviour which is characterized by small errors  for small $\alpha$ and large errors for large $\alpha$ 
is a general phenomenon almost independent from the order chosen.  Particularly  for $\alpha \geq 0.9$ results are discouraging.  

Therefore, to achieve accuracies comparable to standard contour integral methods, we need an additional ingredient — a new, creative idea.

Well then, the solution is a simple association:

For trigonometric functions we have a multiplication formula to reduce the angle $\alpha$ e.g.
\begin{equation}
\label{bnum03_prodTrig}
\sin(\alpha) = 2^{m-1} \prod_{j=0}^{m-1} \sin(\pi j/m + \alpha/m) \qquad m \in \mathbb{N}
\end{equation}
For the gamma function we have the famous multiplication formula from Gauss to reduce the argument:
\begin{equation}
\label{bnum03_prodGamma}
\Gamma(\alpha) = (2 \pi)^{(1-m)/2} m^{m \alpha - 1/2} \prod_{j=0}^{m-1} \Gamma(j/m + \alpha/m) 
\qquad m \in \mathbb{N}
\end{equation}
Let us  search for a similar formula for the parameter $\alpha$ in the Mittag-Leffler function and voila, 
we find  the addition formula ($z \in \mathbb{C}$) from  \cite{erd53} (18.1.24):
\begin{equation}
\label{bnum03_addTalpha}
E_{\alpha, \beta}(z) = {1 \over m}
\sum_{j=0}^{m-1}  E_{\alpha/m, \beta} (z^{\alpha/m} e^{2 \pi i j/m}) \qquad i=\sqrt{-1} \qquad m \in \mathbb{N}
\end{equation}
Until now, it has exclusively been used to obtain results for $\alpha > 1$ or within the range $1 \leq \alpha \leq 2$, as seen in 
\cite{gor02}, \cite{die05}, \cite{hil06}, \cite{sey08}.
\begin{table}
\caption{
Relative error $\max()\rho(x))$ for some parameter sets $\{\alpha, \beta\}$. For reasons of completeness 
$R^{3,2}_{\alpha,\beta}$ values from \cite{sar20} correspond to second order $n=2$  and $R^{7,2}_{\alpha,\beta}$ from \cite{sar20} 
may be compared with fourth order $n=4$.
}
{\begin{tabular}{{rlrr|lr}}
order&  $\alpha=0.9$,    & $\alpha=0.9$,     &$\alpha=0.5$,   &$\alpha=1.0$, \\
   &  $ \beta=1.9$     & $\beta=1.0 $      &$ \beta=1.0$    &$ \beta=1.1$  \\
\hline
\noalign{\smallskip}
2    & 5.52E-02           &  4.88E-01         &  1.84E-02      &   6.39E-01    \\
$R^{3,2}_{\alpha,\beta}$&5.20-E02&4.88E-01     &  -              &   6.39E-01    \\
3    & 1.60E-02           &  2.81E-01         &  1.99E-03      &   5.26E-01    \\
4    & 5.17E-03           &  1.55E-01         &  2.09E-04      &   4.27E-01    \\
$R^{7,2}_{\alpha,\beta}$&5.50-E03&1.64E-01     &  -              &   3.94E-01    \\
5    & 1.72E-03           &  7.81E-02         &  2.14E-05      &   3.40E-01    \\
6    & 5.87E-04           &  3.72E-02         &  2.17E-06      &   2.63E-01    \\
7    & 2.02E-04           &  1.69E-02         &  2.18E-07      &   1.97E-01    \\
8    & 7.01E-05           &  7.42E-03         &  2.17E-08      &   1.43E-01    \\
10    & 8.56E-06           &  1.31E-03         &  2.12E-10      &   6.75E-02    \\
12    & 1.05E-06           &  2.13E-04         &  2.94E-11      &   2.78E-02    \\
\end{tabular}}
\label{bnum203tab1}
\end{table}
Of course, when using the addition formula within the context of the global Pad\'e approximant, 
we need to make the step from negative real argument  $x \in \mathbb{R}^-$ to complex argument $z \in \mathbb{C}$. 

However,  since our derivation is based on the asymptotic expansions (\ref{num203001a}, \ref{num203001b}) which are valid also for 
$|z|\rightarrow 0$ and $|z|\rightarrow \infty$  with the restriction $|\arg(z)| < \pi \alpha$, we only need to  confine 
ourselves to odd $m= 2 r + 1, r\in \mathbb{N}$,  without significant  loss of generality.

We will extend the public method {\tt{mittagLefflerE}} to apply the addition formula if $\alpha > \alpha_{\text{thresh}}$ 
and add a public method {\tt{setThresh}} to adjust the threshold accordingly.

Listing \ref{bnum03_publicX} shows the implementation. 
\begin{listing}
\begin{oframed}
\vskip-3mm
\begin{lstlisting}[
  label={bnum03_publicX},
  style=mystyleCpp,
  language=C++,
  firstnumber=1,
  caption={Extension  GlobalPadeMLF}
]
inline double mittagLefflerE( double alpha_new, double beta_new, double x ) {
...
  double result = 0;
  // introduce addition formula
  if (alpha_new > alphaThresh) {     
    int m = floor(alpha_new / alphaThresh);
    int mm = 2 * m + 1; 
    double mp = 1.0 / mm;
    complex<double> z;
    complex<double> cResult;
     // smaller alpha 
    SetAlphaBetaSolveEqsGetPsQs(alpha_new * mp, beta_new);
     // sum all terms
    for (int k = 0; k < mm; k++) {
      z = pow(x, mp) * exp(2i * PI * mp * (double) k);
      cResult = getPadeApproximant(z,n,n);
      result += cResult.real(); // imags cancel for real x
    }
    result *= mp;
  }                                              
  else {                                    // old version 
    SetAlphaBetaSolveEqsGetPsQs(alpha_new, beta_new);
    result = getPadeApproximant(x,n,n);
  }
  return result;
}
// set threshold
void setThresh(double new_tresh){
alphaThresh = new_tresh;
}
\end{lstlisting}
\vskip-3mm
\end{oframed}
\end{listing}
\index{listings!Extension of in GlobalPadeMLF.hpp}

\begin{figure}[t]
\begin{center}
\includegraphics[ width=\textwidth]{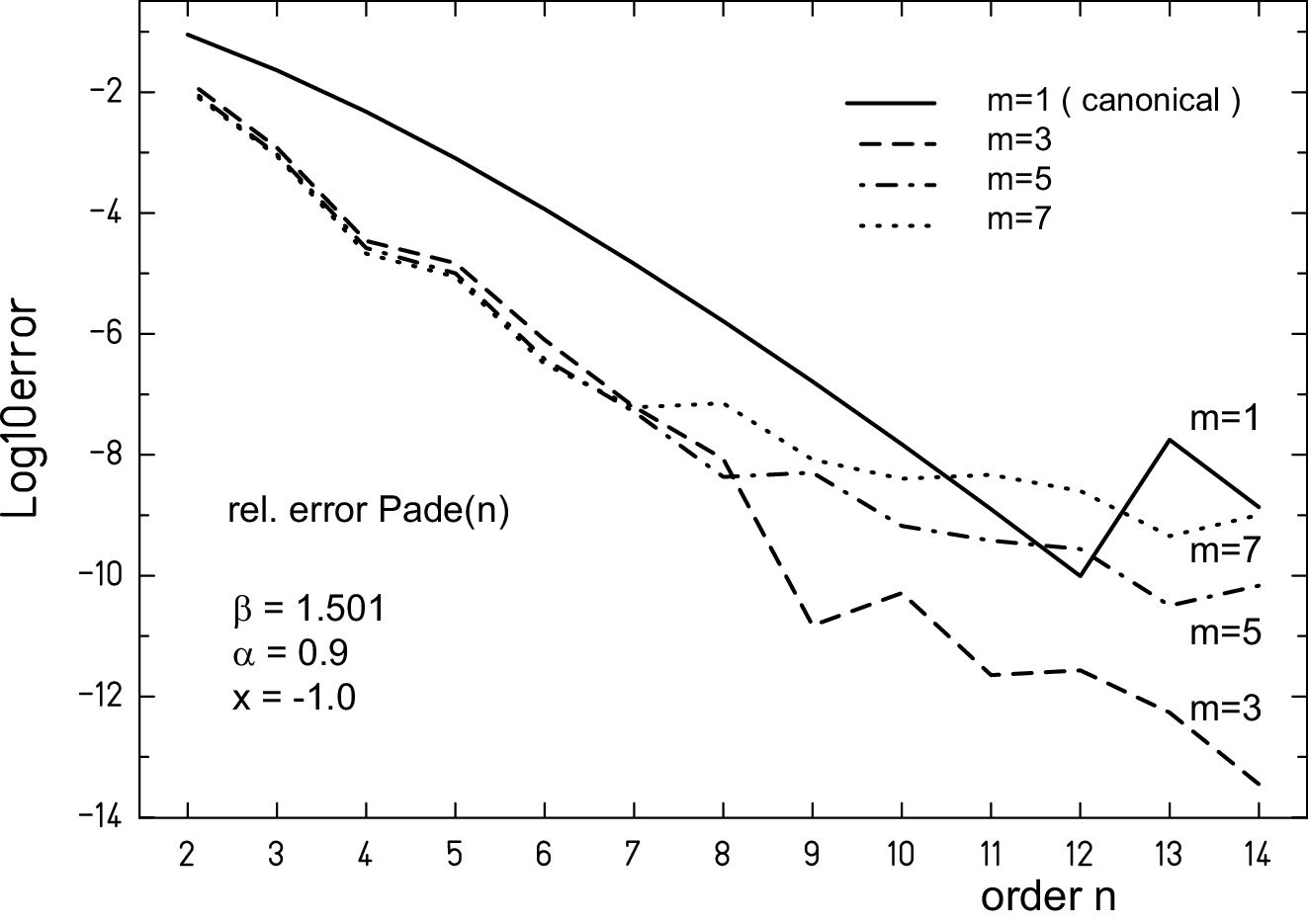}\\
\caption{
\label{bnum03_36}
{ Logarithmic plot for the relative error $\rho(x)$ of 
the Pad\'e approximant of $E_{0.0, 1.5}(-1)$ for increasing order $n=2,...,14$ for different values of $m$ in the 
$m = \lfloor {\alpha \over \alpha_t} \rfloor$ in the addition formula.
$m=1$ corresponds to the classical (no addition formula). The optimum choice is $m=3$ with an accuracy gain of about a factor 100 and Furthermore
leads to accuracies better that 13 digits.
} }
\end{center}
\end{figure}

\begin{itemize}
\item
\ref{bnum03_publicX}:5--20: The member function {\tt{mittagLefflerE}} is extended to check the case
$\alpha > \alpha_{\text{thresh}}$. In that case, the addition formula is applied, leading to 
a new $\alpha/m$.
Argument becomes complex (lines 9--10), bet the result is a real for real original input. 
\ref{bnum03_publicX}:28--30: A member function {\tt{setThresh}} is added to adjust the $\alpha_{\text{thresh}}$. 
\end{itemize}
In Figure \ref{bnum03_36} we plot the relative error $\rho(x)$ for increasing orders $n$ for different   $m = \lfloor {\alpha \over \alpha_t} \rfloor$ in the addition formula.

For $m=3$ we obtain about 2 orders of magnitude in accuracy compared to the classical (no addition formula applied) method. 
For an  accuracy goal of $10^{-6}$ an order of $n\geq 6$ instead of $n \geq 9$ suffices.   
Furthermore we obtain accuracies ($|\rho(x)| < 10^{-13} $) comparable with 
the contour integral methods for a reasonable order $n \geq 14$. 

At this stage, we can already compare these findings with the results obtained for the series expansions 
of the gamma function in Exercises \ref{num0201} and \ref{num0202}. This allows us to make an initial estimate of the 
effectiveness of this method. Since the global Padé approximation is composed of two Horner schemes for the numerator and 
denominator, the complexity and, therefore, the time consumption can be estimated to be approximately double that 
of the gamma function evaluations with comparable accuracy.

In Figure \ref{bnum03_p4567} we compare the relative error $\rho(z)$  of the global Pad\'e approximant 
of order $n=6$  for the Mittag-Leffler function $E_{\alpha=0.9, \beta=1.0}(-z)$ in the 
complex plane for $-3 \leq Re(z), Im(z) \leq +3$.
We consider the cases $m=1$,(no addition formula ) and $m=3$ with addition formula, which leads to an 
evaluation of the Mittag-Leffler function at $\alpha = 0.3 $ instead of $\alpha = 0.9$.

Near the origin ($|z| < 0.1$) the global Pad\'e approximant leads to poor results in both cases ($m=1,3$). 
However, this is also the region where a direct evaluation of the defining series of the Mittag-Leffler function is the best strategy obtaining reliable results.

Besides $z=0$, similarly  poor results are obtain for all $n+1$ zeros of the denominator of the global Pad\'e approximant. For a given order $n$, these zeroes $z_0^k$ are determined by the condition 
\begin{equation}
Q(k,x)=0, \qquad k \in \{0,1,...,n \} 
\end{equation}
Remarkable enough, the set of zeroes (omitting $z_0^0=0$) obtained not using the addition formula lies outside a circle $|z_0^k| > 1.54$, while using the addition formula with $m=3$ leads to $|z_0^k| > 2.2$, which we interpret as an first indication for an 
enhanced stability of results obtained using the addition formula.  

Along the real positive axis, we observe a significant precision enhancement using the addition formula of magnitude $100-1000$ compared to the classical case, reducing the error from about $\log(\rho(m=1,x)) < -2.5$ up to 
$\log()\rho(m=3,x) < -5.0$. 

We conclude that the global Pad\'e approximant used in its standard form 
without applying the addition formula is state of the art but  
leads to poor-quality results especially for large $\alpha$.
\begin{listing}
\begin{oframed}
\vskip-3mm
\begin{lstlisting}[
  label={bnum03_integral},
  style=mystyleCpp,
  language=C++,
  firstnumber=1,
  caption={Contour integral}
]
#include <atomic>  // For counter (thread-safe)
#include <numbers> // C++20: pi
#include <boost/math/special_functions/gamma.hpp>
#include <boost/math/quadrature/gauss_kronrod.hpp>

atomic<long int> counter{0};  // thread save counter 

inline auto K = [](double z, double a, double b,
               atomic<long int> &counter) {
// create the integrand as a lambda function
return [ = , &counter](double t) {
double x = t^2; // coordinate transform x = t^2

double result = pow(x,(1-b)/a)*exp(-pow(x, 1/a))*
(x*sin(numbers::pi*(1-b))-z*sin(numbers::pi*(1-b+a)))/
(x*x - 2*x*z*cos(numbers::pi*a) + z*z);

counter++;         // count function calls
return 2*t*result; // volume element dx->dt=2*t*dt
};
};

inline double MittagLefflerZAB(double z,double a,double b)
{  
// ToDo: specials, only one example implemented
if( almostEqual(z, 0.0) ){ 
return 1.0 / boost::math::tgamma(b);
}  
// prepare input for quadrature
auto integrand = K(z,a,b,counter);
// define the interval for the transformed variable t
double lowerLimit = 0.0; 
double upperLimit = numeric_limits<double>::infinity();
// prepare precision attributes 
int numNodes = 15;             // number of nodes
double errorTolerance = 1.e-6; // error tolerance 
int adaptLevel = 10;           // iterations 
// central function call
double Q = boost::math::quadrature::
gauss_kronrod<double, numNodes>::integrate(integrand,
lowerLimit, upperLimit, adaptLevel, errorTolerance);

return Q / (a * numbers::pi);
}
\end{lstlisting}
\vskip-3mm
\end{oframed}
\end{listing}
\index{listings!Gauss-Kronrod quadrature}

However using the addition formula with $m=3$ in combination with the global Pad\'e approximant provides a path to 
reliable, high-precision results, at least for the case $z \in \mathbb{R}^-$. This approach appears to be a promising alternative to the classical contour integral evaluation
used up to now in the standard library implementations. 

\section{Comparison to the canonical quadrature of the defining contour integral   }

To get an idea of the expected performance gain of this new method we will compare the efficiency of our solution with an
standard implementation of the contour integral evaluation method for the simplified case $0 < \alpha < 1, \beta < 1 + \alpha$. 
In this case, we have to evaluate the integral \cite{die05}:
\begin{eqnarray}
\label{bnum203_GK}
E_{\alpha, \beta}(z)
&=&
\int_0^\infty  K(\alpha,\beta,\xi,z) d\xi
\end{eqnarray}
with      
\begin{eqnarray}
\label{bnum203_K}
K(\alpha,\beta,\xi,z)
&=&
{ \xi^{(1-\beta)/a} e^{-\xi^{1/\alpha}}
\over
\alpha \pi
}
{\xi \sin(\pi(1-\beta)) - z \sin(\pi(1-\beta+\alpha))
\over
\xi^2 - 2 \xi z \cos(\alpha \pi) + z^2
}  \nonumber \\
\end{eqnarray}
Since there is no analytic solution, we will use the Gauss-Kronrod quadrature, which allows for an efficient
error estimation and is available in nearly all libraries dealing with numerical integration.

We will employ the methods available from the {\tt{boost}} template library.

Listing \ref{bnum03_integral} shows our implementation, which defines two functions the integral procedure
{\tt{MittagLefflerZAB}} and the integrand {\tt{K}}: 
\begin{itemize}
\item
\ref{bnum03_integral}:8-21: The integrand $K(\alpha,\beta,x,z)$ from (\ref{bnum203_K}) is realized as a lambda function (anonymous). 
We added a thread-safe counter 
to obtain information on the number of function calls in single- as well as multi-thread environments.

Then,  in lines \ref{bnum03_integral}:12,19: we added a coordinate transform of type $x = t^s$ with volume element $dx = dt s t^{s-1}$. This transformation allows us to influence the initial distribution of abscissas. Numerical experiments show
that choosing $s=2$ reduces the number of function calls required to achieve a given accuracy goal by almost a factor 2, leading 
to a corresponding performance gain.
\item
\ref{bnum03_integral}:22--42: The method {\tt{MittagLefflerZAB}} prepares and calculates the integral calling the relevant library
function {\tt{ boost::math::quadrature

::gauss\_kronrod<double, numNodes>::integrate}} in line 37.
\item
\ref{bnum03_integral}:24--26: In a first step, special cases are checked. Here we show as an example the case $z=0$. We have omitted additional
necessary checks like not yet implemented cases e.g. $\alpha > 2$,  $\alpha = 0$ or $\beta > 1 + \alpha$. A good example for a complete
treatment of special cases is in \cite{die05}.
\item
\ref{bnum03_integral}:30--31: It is no problem to define the integration limits $\{0, \infty \}$.
\item
\ref{bnum03_integral}:33--35: Setting the error handling is one advantage of the Gauss-Kronrod quadrature method. The initial number of nodes
is preferable one of 

\noindent
$\{15,31,41,51,61\}$, since the corresponding abscissa and weights are precomputed, so there is no overhead.
Remarkable enough, these are standard values, which are common to nearly all implementations available (NAG, GSL, partly R...). Only the 
GLS-library offers one more ({\tt{numNodes = 21}}) option, see also \cite{kro64}, \cite{kro65}, \cite{not16}. We define  a error maximum 
{\tt{errorTolerance}} and a maximum {\tt{adaptLevel}} of iteration steps, each generating about $2 numNodes + 1$  function calls per step.    
\end{itemize}
We now may compare the precision and performance of the quadrature versus Pad\'e approximant method to obtain high accuracy values 
for the generalized Mittag-Leffler function $E_{\alpha, \beta,}(-x)$.

\begin{figure}[t]
\begin{center}
\includegraphics[width=\textwidth]{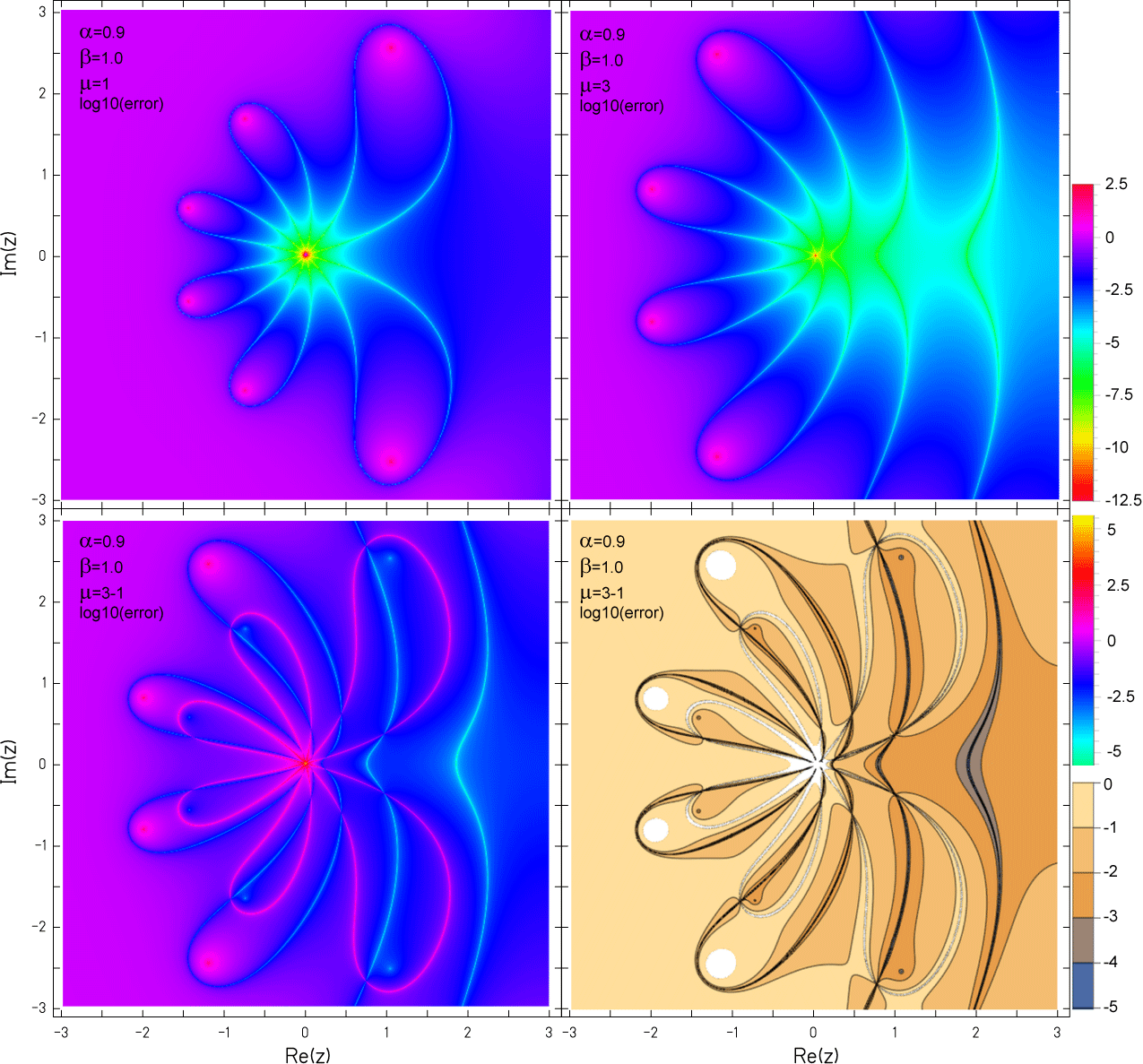}\\
\caption{
\label{bnum03_p4567}
{ Plots of the relative error of the global Pad\'e approximant for the Mittag-Leffler function $E_{\alpha=0.9, \beta=1.0}(-z)$ for complex argument z with $-3 \leq Re(z), Im(z) \leq +3$.
Upper left: classical error (no addition formula applied). Upper right: error with addition formula ($m=3$) applied. Lower left: error difference (negative values are better). Lower right: same as lower left, but contours.        
} }
\end{center}
\end{figure}
\begin{table}
\caption{
A comparison of Pad\'e approximant with increasing order $n$ to the Gauss-Kronrod qwuadrature with increasing error tolerance {\it{tol}}
solving integral (\ref{bnum203_int2}) for $\alpha=0.9,\beta=1.0,s=4$.
From left to right columns show order n, relative error $\tilde{\rho}(m=1)$ for standard Pad\'e approximant (no addition formula applied), 
total time elapsed in milliseconds ms, relative error $\tilde{\rho}(m=3)$ for Pad\'e approximant with addition formula applied, 
total time elapsed in milliseconds ms. This is followed by the listing of predefined maximum error $10^{\it{tol}}$, resulting error $\tilde{\rho}$
elapsed time for the Gauss-Kronrod quadrature of (\ref{bnum203_GK}) and number of function calls {\tt{counts}}. The symbol \# indicates 
results dominated by rounding errors and therefore invalid.   
}
{\begin{tabular}{{rlr|lr||rrlrr}}
order $n$&  $\tilde{\rho}(m=1)$   &  ms     &$\tilde{\rho}(m=3)$  & ms  & tol & $\tilde{\rho}$ &ms & counts\\
\hline
\noalign{\smallskip}
   &            &          &                &           &  0&6.5E-03&  10& 15\\
1    & 1.55E-00   &  0.09    &  1.91E-01      &   0.76    & -1&2.4E-02&  65& 62\\
2    & 2.90E-01   &  0.09    &  1.87E-02      &   0.97    & -2&2.3E-03& 101& 98\\
3    & 5.99E-02   &  0.10    &  5.48E-04      &   0.97    & -3&2.0E-04& 146& 132\\
4    & 1.05E-02   &  0.11    &  1.22E-04      &   1.10    & -4&5.2E-05& 158& 158\\
5    & 1.55E-03   &  0.13    &  1.45E-05      &   1.34    & -5&3.7E-06& 194& 194\\
6    & 1.99E-04   &  0.14    &  8.24E-08      &   1.29    & -6&5.6E-12& 250& 236\\
7    & 2.26E-05   &  0.16    &  9.31E-08      &   1.40    & -7&1.1E-12& 286& 273\\
8    & 2.31E-06   &  0.18    &  5.04E-09      &   1.58    & -8&1.2E-12& 330& 314\\
9    & 2.16E-07   &  0.19    &  3.15E-10      &   1.76    & -9&7.6E-15& 422& 398\\
10    & 1.85E-08   &  0.21    &  7.89E-11      &   1.77    &-10&4.0E-15& 514& 485\\
11    & 1.47E-09   &  0.27    &  1.20E-12      &   2.50    &-11& -     & -  & -  \\
12    & 1.08E-10   &  0.25    &  9.26E-13      &   2.08    &&&& \\
13    & \#2.34E-08   &  0.27    &  2.58E-13      &   2.15    &&&& \\
14    & \#1.57E-09   &  0.36    & \#6.04E-13      &   2.45    &&&& \\
15    & \#2.99E-06   &  0.32    & \#1.88E-12      &   2.64    &&&& \\
\end{tabular}}
\label{bnum203tabInt}
\end{table}
In Table \ref{bnum203tabInt} we list results of the numerical approximation of the integral  
\begin{equation}
\label{bnum203_int2}
\int_0^1 dx E_{\alpha, \beta}(-x) \approx 10^{-s} \sum_{i=0}^{10^s} \tilde{E}_{\alpha, \beta}(-10^{-s} i) + \tilde{\rho} 
\end{equation}
for $\alpha=0.9,\beta=1.0,s=4$, which gives a nice estimate for the relative error $\tilde{\rho}$ of the two 
approximations $\tilde{E}$ in a practical application. 

The standard Pad\'e approximant produces reliable results with increasing accuracy up to $\tilde{\rho} \approx 10^{-9}$ for orders up to  $n\approx 12$.
When the addition formula with $m=3$ is enabled, reliable results  
with increasing accuracy are obtained up to $\tilde{\rho}  \approx  10^{-13}$ for orders up to  $n\approx 13$, but this method requires about 10 times more CPU time.

The Gauss-Kronrod quadrature  produces reliable results with increasing accuracy up to $\tilde{\rho} \approx 10^{-15}$ for a tolerance level  $tol\approx 10^{-10}$,
but it requires $\approx 1600$ times more CPU time compared to the standard Pad\'e approximant and about  $\approx 200$ times more CPU time compared to the Pad\'e approximant
with the addition formula.

Therefore, the optimal method depends on the required accuracy.

As a rough guideline: 
\begin{itemize}
\item
For {\tt{float}}-type precision 
($\tilde{\rho} \leq 10^{-6}$)
use the standard Pad\'e approximant and disable the addition formula implementation.
\item
For $\tilde{\rho}$ in the range $10^{-7} \leq \tilde{\rho} \leq 10^{-12}$ enable the addition formula implementation. 
\item
To achieve the highest possible accuracy with double-type precision  ($\tilde{\rho} \leq 10^{-12}$) solve the defining contour integral with an appropriately chosen path 
with the help of the Gauss-Kronrod quadrature.   
\end{itemize}

\begin{figure}[t]
\begin{center}
\includegraphics[width=\textwidth]{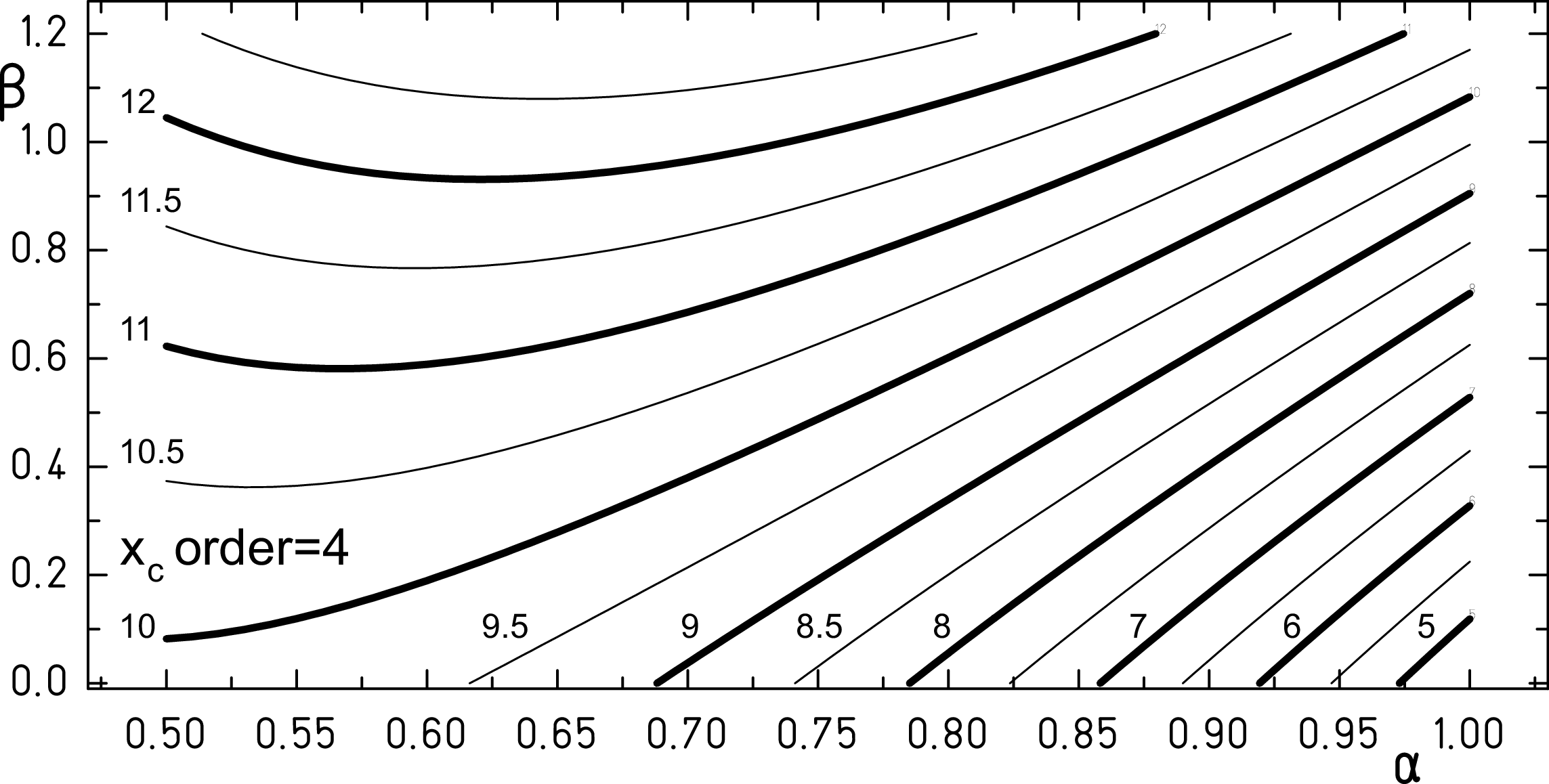}\\
\caption{
\label{bnum03_p30202}
{ Critical length $x_c^\infty$ determined by the condition  $\forall x > x_c^\infty, \rho(x, m=1) < \rho(x, m=3)$ holds.        
} }
\end{center}
\end{figure}

An additional remark on the error behaviour as a function of x: 

From Figure  \ref{bnum03_p4567} we can deduce, that for very small $|x|$
the standard Pad\'e approximant without the  addition formula implementation shows better error behaviour,  $\forall x < x_c^0, \rho(z, m=1) < \rho(z, m=3)$? 
In this region,  the truncated series expansion (\ref{num203001a}) for $E_{\alpha,\beta}(z)$ could be the best choice for an approximation. 

The question arises
whether the same holds true for very large $x \gg 1$. 
Is there a critical length $x_c^\infty$ such that  
$\forall x > x_c^\infty, \rho(x, m=1) < \rho(x, m=3)$?
This is indeed the case.

In Figure \ref{bnum03_p30202} we present a plot of  $x_c^\infty(\alpha, \beta)$ for order $n=4$ of the    Pad\'e approximant.
A least-squares fit with a polynomial up to second order in $(\alpha,\beta)$ yields:
\begin{eqnarray}
\label{bnum203_xcrit}
x_c^\infty(\alpha, \beta, n=4)  &=& 5.36 + 18.94 \alpha -  19.85 \alpha^2 - 1.59 \beta + 6.13 \alpha \beta +  0.53 \beta^2  \nonumber \\
&& 
\end{eqnarray}
An extended version of the addition formula implementation should test this condition. Furthermore, the asymptotic expansion 
for $x \rightarrow \infty $ (\ref{num203001b}) should be considered for large arguments.

\section{Conclusion}
We presented a combination of arbitrary order Pade approximation method with the addition formula for the numeric implementation 
of the Mittag-Leffler function for the case $\alpha<1$  and compared the efficiency with the standard numerical quadrature methods of the defining contour integral.    

Our tests lead to the conclusion, that the  optimal method depends on the required accuracy.
\begin{itemize}
\item
For {\tt{float}}-type precision  ($\tilde{\rho} \leq 10^{-6}$) use the standard Pad\'e approximant and disable the addition formula implementation. 
This method is about a factor $\approx$ 1500 compared to Gauss-Kronrod quadrature.

For a relative error  in the range $10^{-7} \leq \tilde{\rho} \leq 10^{-12}$ enable the addition formula. 
Using this strategy enhances accuracy significantly a factor greater than 100 for large $\alpha$. 
However, due to the step from real to complex argument it takes a factor 8-10 more than the standard approach omitting the addition formula.
Therefore this method is about a factor $\approx$ 150 compared to Gauss-Kronrod quadrature.

To achieve the highest possible accuracy with double-type precision  ($\tilde{\rho} \leq 10^{-12}$) 
or if time and money does not play any role 
solve the defining contour integral with an appropriately chosen path 
with the help of the Gauss-Kronrod quadrature.   
\end{itemize}

\begin{acknowledgements}
We thank A. Friedrich for useful discussions and suggestions.
\end{acknowledgements}




\bigskip  

\small 
\noindent
{\bf Publisher's Note}
Springer Nature remains neutral with regard to jurisdictional claims in published maps and institutional affiliations.

\end{document}